# A Pilot Study on Mandarin Chinese Cued Speech

Li Liu and Gang Feng

Cued Speech (CS) is a communication system developed for deaf people, which exploits hand cues to complement speechreading at the phonetic level. Currently, it is estimated that CS has been adapted to over 60 languages; however, no official CS system is available for Mandarin Chinese. This article proposes a novel and efficient Mandarin Chinese CS system, satisfying the main criterion that the hand coding constitutes a complement to the lips' movements. We propose to code vowels [i, u, y] as semiconsonants when they are followed by other Mandarin finals, which reduces the number of Mandarin finals to be coded from 36 to 16. We establish a coherent similarity between Mandarin Chinese and French vowels for the remaining 16 finals/vowels, which allows us to take advantage of the French CS system. Furthermore, by investigating the lips viseme distribution based on a new corpus, an optimal allocation of the 16 Mandarin vowels to different hand positions is obtained. A Gaussian classifier was used to evaluate the average separability of different allocated vowel groups, which gives 92.08%, 92.33%, and 92.73% for the three speakers, respectively. The consonants are mainly designed according to their similarities with the French CS system, as well as some considerations on the special Mandarin consonants. In our system, the tones of Mandarin are coded with head movements.

Keywords: Cued Speech, Mandarin Chinese Cued Speech, communication for the deaf community

Communication is one of the most important parts of human life, and more and more attention has been paid to improving the communication among people with disabilities. It was reported by the World Health Organization that more than 5% of the world's population (466 million people) has a hearing loss (432 million adults and 34 million children). Speechreading is one of the more common communication methods for many deaf people. It helps individuals who are deaf or hard of hearing access spoken language (Dodd & Campbell, 1987; Woodward & Barber, 1960).

However, there still exists a problem with speechreading because of the similarity of labial shapes, such as the ambiguity of vowels [y] and [u]. As a result, this problem makes it difficult for individuals who are deaf or hard of hearing to access spoken language only by the traditional oral education. Many methods have been proposed to overcome this problem, and most of them use hand coding to provide additional information.

Cued Speech was invented by R. Orin Cornett (1967) at Gallaudet University to enable easier access to spoken language.

Li Liu received her doctoral degree in 2018 from Gipsa-lab, University Grenoble Alpes, Grenoble, France, and she currently works as a research scientist in Shenzhen Research Institute of Big Data, The Chinese University of Hong Kong, Shenzhen, China. Gang Feng received a doctoral degree from Grenoble Institute of Technology, France, in 1986 and he is a professor at Gipsa-lab, University Grenoble Alpes, Grenoble, France.

In this system, a special hand coding (i.e., a combination of different handshapes and positions near the face) complements the speechreading process to enhance speech perception. More precisely, the handshapes are used on one side of the face and neck to code consonants, while on the other side the hand positions code vowels. An example of the French CS is shown in Figure 1,[1] where 5 hand positions (i.e., side, mouth, neck, cheek, and chin) are used to code vowels and 8 different handshapes are used to code consonants.

Around the world, CS is becoming more and more popular as an aid to improve the communication between deaf or hard of hearing children and their hearing family members, and has been adapted to more than 60 languages. The National CS Association (NCSA)[2] for American English CS, the Cued Speech Association United Kingdom (CSAUK)[3] for British English CS and the Association nationale pour la Langue française Parlée Complétée (ALPC)[4] for Langue Francaise Parlée Complétée (LPC) have been established to generalize this system. In previous research (Nicholls & Mcgill, 1982; Alegria, Dejean, Capouillez, & Leybaert, 1990), it has been shown that when individuals who are deaf or hard of hearing use CS, they can almost completely access spoken language.

Destombes (1982) demonstrated that a deaf child was able to acquire a complete and accurate model of spoken language using visual information only. Périer and De Temmerman (1987) suggested that CS can reduce the barriers that deaf children experience in their initial communication with their hearing parents. Leybaert and Charlier (1996) showed that the CS system allows deaf individuals to have a complete phonological representation of a language. By comparing their performances with hearing people in reading and writing abilities, Leybaert and Charlier. also showed that CS can help them develop these abilities.

Several studies have been conducted on different versions of CS that show how the system can help speech perception for deaf or hard of hearing individuals by complementing hand coding with speechreading. For American English CS, readers may refer to the following studies: Ling and Clarke (1975); Clarke and Ling (1976);

**Figure 1.** French CS (LPC) system

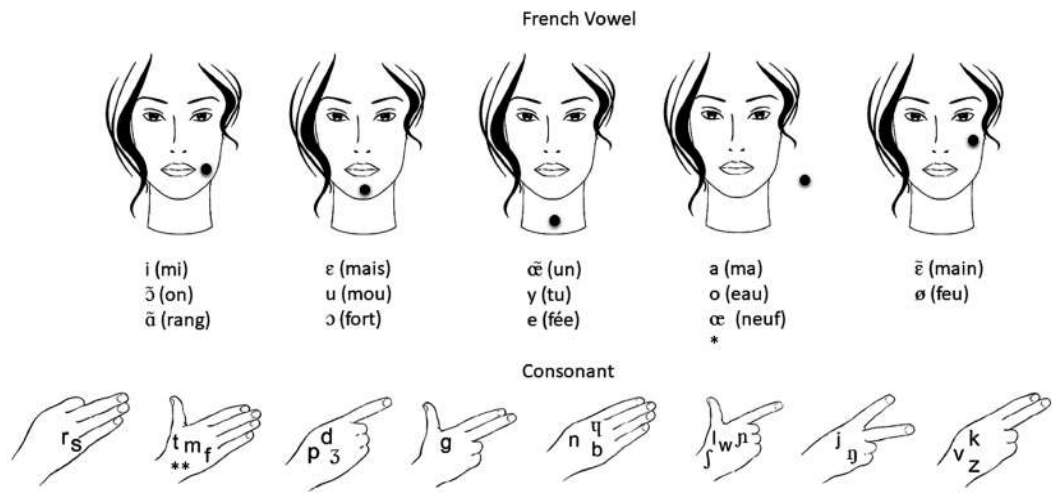

Nicholls and McGill (1982); Uchanski et al. (1994); and Charlier, Hage, Alegria, and Périer (1990). Concerning the French CS system, readers may be interested in Alegria, Charlier, and Mattys (1999); Attina, Beautemps, and Cathiard (2002, 2004); Attina (2005); Aboutabit (2007); Liu (2018); Liu Hueber, Feng, and Beautemps (2018); Liu, Li, Feng, and Zhang (2019); and Heracleous, Aboutabit, and Beautemps (2009).

In the following two studies, we can see how American English and French CS systems produce dramatic improvements to children's speechreading ability. When studying American English CS, Nicholls and McGill (1982) tested reception of multiple messages in seven different environments by 18 hard of hearing children aged 9 to 16 years. The experiment showed that percentages of correct perception of English syllables increased from 30% using speechreading only to 83.5% using CS (with 53.5% improvements). As for the French CS system, the work of Alegria et al. (1999) tested the perception of French signal words in 7 "CS-early" children, aged 8 to 12 years, and 24 "CS-late" children, aged 11 to 19 years. CS-early children were exposed to CS before the age of 2, and CS-late children were exposed to CS after the age of 2. They found that the correct perception of French words in the CS-early children increased from 40% using speechreading only to 77% using CS, and perception increased from 50% to 70% for the CS-late children. These works demonstrate that American English and French CS systems largely enhance speechreading by combing lip and hand cues.

Most CS versions of spoken languages (i.e., British English, French, and Dutch) originate from the first work of Dr. Cornett (1994). As far as we can determine, there is no official research work or available CS system for Mandarin Chinese. However, according to the China Disabled Persons' Federation,[5] about 21 million people have hearing loss out of the 60 million people with disabilities in China. For the moment, Chinese Sign Language (Fischer & Gong, 2010; Yang & Fischer, 2002; Chena, Gao, Fang, Wang, & Yang, 2003), which has been developed since the late 1950s, is the most popular communication method among Chinese deaf people. One of the reasons that Mandarin Chinese CS is nonexistent probably lies with the great complexity of Mandarin Chinese. In fact, according to the NCSA, a Mandarin Chinese CS system is difficult to build due to the large number and complicated combinations of Mandarin phonemes. Besides, some confusion between phonetic pronunciation and Pinyin makes the development of a Mandarin Chinese CS system very difficult.

This work aims at proposing a new Mandarin Chinese CS system that satisfies the main criterion (Cornett, 1994) that hand and lips coding are complementary (i.e., the phonemes with similar lipshapes should be distinguished by different hand coding). Moreover, the proposed system has been optimized so that CS cuers need to use minimum energy to code it, and the interlocutors can easily perceive the cuer's gestures. Mandarin Chinese CS is not invented to replace Chinese Sign Language but to make individuals who are deaf or hard of hearing (especially children) have easier access to the spoken language. And it allows them another option to learn Mandarin.

In summary, this work contains the following four contributions in order to build our Mandarin Chinese CS system:

1. All the compound finals starting with *i* [i], *u* [u], *ü* [y][6] are coded using

semiconsonants [j], [w], [ɥ]. This strategy reduces the number of Mandarin finals to be coded by hand positions from 36 to 16, and is much more efficient than coding them by two or three successive simple vowels.

2. The 16 remaining finals/vowels can be coded by 5 hand positions. In order to optimize the allocation of these finals/vowels into 5 hand positions, two studies were carried out. The first one exploits a similarity between Mandarin and French vowels. This similarity allows us to establish a first vowel allocation, which possesses a certain optimality.[7] Secondly, lips parameter distributions for different vowels coded by hand positions are studied based on a new corpus recorded specifically for this work, which permits optimization of the vowel allocation. Compared with the British/America English CS, a significant advantage of the proposed method is that no Mandarin compound finals are coded by hand slides (i.e., hand movements from one position to another), which makes our Mandarin Chinese CS system much simpler and more efficient.

3. The most commonly used consonants are distributed in the same manner as French consonants, which has already confirmed its optimal performance. For the special consonants in Mandarin Chinese, their distribution is determined to distinguish their corresponding lipshapes. In addition, the consonants that can be combined with semiconsonants [j], [w], [ɥ] are not allocated to the same hand shape group.

4. We propose to use four head movements to code the Mandarin Chinese tones. In this way, the cuer's hand is able to be wholly focused on the phoneme coding.

## Coding of Mandarin Finals: Methodology

Mandarin Chinese is organized around syllables that are created by an initial followed by a final. There are 35 finals (when *er* is considered, this number becomes 36) and 21 initials (Manser Ren, Wu, & Zhu, 2003). Initials are almost all consonants, while finals can be divided into two categories: simple finals and compound finals (Manser et al., 2003). Simple finals contains 6 single vowels (i.e., *a, o, e, i, u, ü*, see the elements with a gray background in Table 1). Compound finals can be diphthongs (e.g., *ai, ei, ao, ou*), nasalized vowels (e.g., *an, eng*, etc.), or finals that begin with *i, u, ü*, followed by a single vowel, diphthong, or a nasalized vowel, for example, *ia, uai, üan* (see the elements inside the right rectangle box in Table 1).

**Table 1.** Mandarin finals table (in Pinyin) (Manser et. al., 2003)

| Mandarin finals | *i* | *u* | *ü* |
|---|---|---|---|
| *a* | ia | ua | |
| *o* | | uo | |
| *e* | ie | | üe |
| ai | | uai | |
| ei | | uei | |
| ao | iao | | |
| ou | iou | | |
| an | ian | uan | üan |
| en | in | uen | ün |
| ang | iang | uang | |
| eng | ing | ueng | |
| ong | iong | | |

*Note.* The six symbols with gray backgrounds are simple finals/single vowels. All the other finals are compound finals. The symbols inside the left rectangle box are diphthongs and nasalized vowels. The symbols inside the right rectangle box are compound finals starting with *i, u,* and *ü*. Note that the vowel *er* is not included in this table.

Due to the large number (36) of Mandarin finals, we can imagine a great complexity to distribute them in CS system only by several hand positions compared with the British/America English systems.[8] According to Cornett (1967), people can relatively easily distinguish three different lip configurations (open, flattened-relaxed, and rounded), thus one hand position could be used to code three different vowels that have distinguishable lips forms. In this case, about 12 hand positions should be needed to code 36 Mandarin finals, which is not reasonable.

Another way to code compound finals may be the combination of two or three successive vowels. This is very similar with the way of coding diphthongs in the English CS system, which uses slides between different hand positions (Nicholls & Mcgill, 1982; Alegria et al., 1990). However, due to the large number of compound finals in Mandarin, too many hand slides would be necessary. This would make the coding procedure very time-consuming, and complicated for the cuers to learn and perceive.

## Coding of Compound Finals Starting With *i*, *u*, *ü* Using Semiconsonants

Due to the fact that the high vowels *i*, *u*, *ü* at the beginning of a compound final can be seen as a glide (Lin, 2007), we propose to code them in Mandarin Chinese CS by semiconsonants [j], [w], [ɥ] instead of vowels *i*, *u*, *ü*. In this way, [j], [w], [ɥ] are coded by handshapes instead of hand positions, and thus avoid the use of hand slides. In fact, the change of handshapes during the trajectory from one hand position to another can be realized easier and more rapidly than the change of hand positions (Cornett, 1967; Nicholls & Mcgill, 1982). Moreover, [j], [w], [ɥ] combined with a preceding consonant form a consonant group. The decoding of such consonant groups is easier and less ambiguous in the case of changing hand shapes than changing hand positions. Moreover, the optimization of the consonant allocation using handshapes is comparably easy to avoid confusion of these semiconsonants with other consonants.

Recall that CS systems are phonetic-based instead of orthographic-based. Thus, in its orthographic form, when a compound final is used alone without a consonant ahead, *i*, *u*, *ü* are changed to *y*, *w*, *yu*, respectively. But if there is a consonant before the compound final, its orthographic form remains unchanged. For example, for *uai*, if there is no consonant ahead, it will be written as *wai*. However, if there is a consonant ahead, such as *kuai*, *uai* will keep its original form instead of changing to *wai*. This is a key difference compared with our proposed method for Mandarin compound finals. In fact, in all cases, a compound final starting with *i*, *u*, *ü* is always coded using a semiconsonant, no matter if this compound final is alone or precedes a consonant. This is coherent with the fact that *uai* and *wai* have the same pronunciation.

In addition, this coding approach is consistent with our educational methods of Mandarin Chinese. For example, *liang* is spelled as *l* followed by *iang*, but pronounced as *l* [l] followed by *yang* [yɑŋ]. In our proposed Mandarin Chinese CS system, *liang* is coded in the exact same way (i.e., *l* [l] followed by *yang* [yɑŋ]).

By using semiconsonants to code the compound finals starting with *i*, *u*, *ü*, the number of the remaining finals that need to be distributed to different hand positions is considerably reduced (from 36 to 16). These finals (see Table 1) contain 6 single vowels, 4 diphthongs, 5 nasalized vowels, and also *er*. For the sake of simplicity, in the remainder of this article, we will call them "**vowels**" in Mandarin.

## Coding of the 16 Mandarin Vowels

Recall that two main criteria should be considered when proposing a new CS system: (1) vowels that have similar lipshapes should be distributed into different hand position groups, and (2) the minimum number of hand position groups should be used to reduce the amount of energy needed.

The French CS system is an optimized CS system that follows that criteria. ALPC has presented strong and compelling evidence that when using LPC, French deaf children perform better in school and their hearing parents can communicate better with their deaf children at home. Moreover, the effectiveness of LPC has been shown in the literature (Charlier et al., 1990; Alegria et al., 1999; Attina et al., 2002; Attina et al., 2004; Attina, 2005; Aboutabit, 2007; Heracleous et al., 2009).

We remark that in LPC (LaSasso et al., 2010), 16 French vowels are distributed in 5 hand positions without any hand slides. In this section, we explore the similarity between the 16 Mandarin vowels and the French vowels. This similarity allows us to build a preliminary allocation of the Mandarin vowels. This preliminary allocation will then be optimized by studying the lip parameter distribution for the Mandarin Chinese. It should be mentioned that our work does not constitute a detailed phonetic phonological study concerning these two languages. The similarity is just exploited to determine a first vowel allocation that will then be optimized.

### Similarities Between Mandarin and French Vowels

Comparing French and Mandarin vowels, evidently, there is no simple one-to-one mapping since they are different in nature. In fact, Mandarin vowels are a bit more complicated than French vowels because all 16 French vowels are monophthongs, while for Mandarin, 6 of them are monophthongs, 4 are diphthongs, and 5 are nasalized vowels. However, we still find that they have some remarkable similarities (see Fig. 2).

**Figure 2.** Similarity between Mandarin and French oral vowels, on the vowel space. Mandarin vowels are in left and French vowels in right. The flash represents the evolution tendency of Mandarin diphthongs. The black curve pointing to *o* [o] means that it is often used with *u* [u]. As for the nasalized vowels, the correspondences are shown in the bottom table.

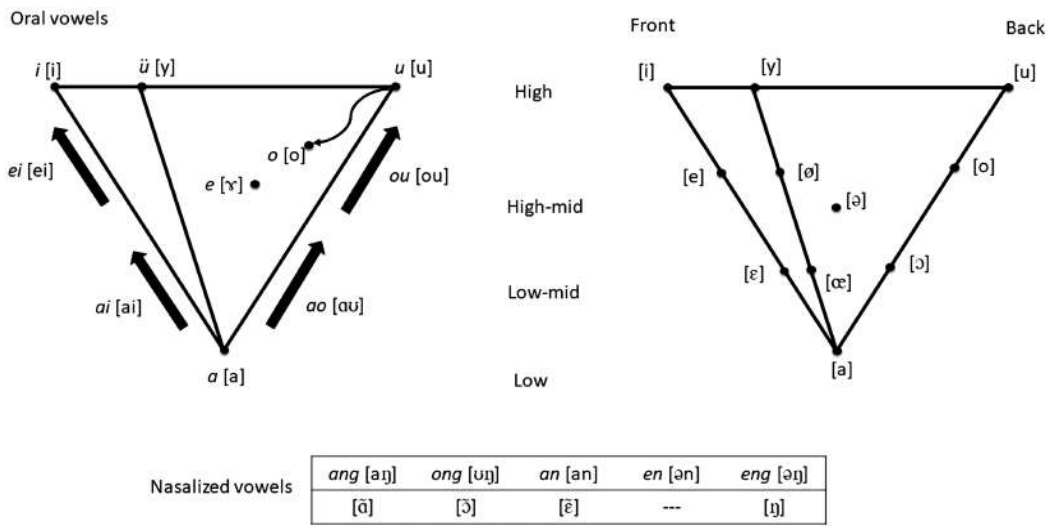

Similarities Concerning Monophthongs

The Mandarin monophthong concerns the following 6 vowels: *a* [a], *o* [o], *e* [ɤ], *i* [i], *u* [u], and *ü* [y]. Based on the phonetic and phonological knowledge of the French and Mandarin, we can formulate the following remarks.

1. Among them, four cardinal vowels *a* [a], *i* [i], *u* [u], *ü* [y] correspond to French vowels [a], [i], [u], [y].
2. It should be noted that the Mandarin vowel *i* has some variations in pronunciation. More precisely, after the consonants *j*, *q*, *x*, etc., the vowel *i* is really pronounced as [i]. However, after the consonants *z*, *c*, *s*, the vowel *i* is pronounced as [ɿ], and after *zh*, *ch*, *sh*, pronounced as [ʅ].[9] This phenomenon causes *i* to have different lip shapes when it is behind different consonants. This point will be discussed and considered when optimizing CS vowel allocations in the next section.
3. Concerning the back unrounded mid-vowel *e*, it is most used phonetic transcription is [ɤ], which is the allophone of the neutral vowel [ə] (Duanmu, 2007). It can be considered to be similar as the neutral vowel [ə] in French.
4. The Mandarin vowel *o* does not correspond exactly to French vowel [o], even though it is often transcripted to [o]. In fact, *o* is rarely used alone, and is almost always used after *u* to form diphthong *uo*. More precisely, *o* is just a shortened form of *uo* when it follows *b*, *p*, *m*, *f* (only one exception concerning the word *lo*) (Duanmu, 2007).

Similarities Concerning Diphthongs and Nasalized Vowels

First, we examine four diphthongs: *ai* [ai], *ei* [ei], *ao* [au], and *ou* [ou] are considered. Strictly speaking, these diphthongs cannot be described only by a fixed position on the vowel space, but by a transition from one location to another (Lin, 2007). However, when examining the evolution patterns of these Mandarin diphthongs on the vowel space, we observe that their trajectories pass near some French vowels. More precisely, *ei* [ei], *ai* [ai], *ou* [ou], and *ao* [au] are close to French vowels [e], [ɛ], [o] and [ɔ], respectively. Indeed, these four vowels constitute four important positions on the vowel space: front mid-close, front mid-open, back mid-close, and back mid-open. Although these four Mandarin vowels are diphthongs, previous literature has reported that they tend to be weakened and monophthongized (Dow, 1972).

Then, we consider the nasalized vowels *an*, *en*, *ang*, *ong*, and *eng*. The goal is to establish an approximate correspondence between Mandarin Chinese nasalized vowels and French nasal vowels, without evoking the delicate differences between a nasalized vowel and a nasal vowel (Feng & Castelli, 1996). We observe that *an*, *ang,* and *ong* are very close to [ɛ̃], [ã], and [õ], respectively. However, there is no evident correspondence of *en* to a known French nasal vowel (see Figure 2). As for *eng*, it corresponds to the final consonant [ŋ].

*Preliminary Vowel Allocation*

Based on the previously mentioned similarities with French vowels, a primary vowel allocation of 11 Mandarin vowels to five hand positions is first established (see Table 2). Then, for the remaining vowels that do not correspond to the French vowels, we categorized them by the following criterion: The vowels with similar lip shapes should be arranged into different hand position categories. Based on the knowledge that *eng* has quite a different lip shape than *ü* and *ei*, we decided to put *eng* in P5 (see Table 2). Based on similar reasons, we distributed *en* in P2, considering

**Table 2.** Allocation of 11 of the 16 Mandarin vowels (in Pinyin) based on the similarities with French CS

| Hand Positions | | Vowels |
|---|---|---|
| cheek | P1 | *an* |
| side | P2 | *a, ou* |
| mouth | P3 | *i, ong, ang* |
| chin | P4 | *ai, u, ao* |
| neck | P5 | *ü, ei* |

that *en* is rather different with *a* and *ou* in terms of lip shape.

After we allocated the vowels discussed previously, we noticed that in position P1, there is only one vowel: *an*. *an* has a different lip shape than *e* and *o*. Thus, these three vowels can be allocated to P1. One may think that *e* and *o* have similar lip shapes and may be confused if they are in the same hand position. However, we note that *e* and *o* almost never occur after the same consonants. More precisely, *o* only follows consonants *b*, *p*, *m*, *f* or naturally forms *uo* (or *wo*) with *u*. However, this is not possible for *e* since it can only follow other consonants (except for the one exception *me*). Thus, the consonant that precedes *e* or *o* prevents any confusion by performing different handshapes.

Finally, we consider *er*, which appears rarely in Mandarin. After the previous steps, 15 Mandarin vowels (except *er*) are distributed to five hand positions (three vowels per one position). *er* should be allocated to a hand position with minimal confusion with other vowels. We found that P2, P4, and P5 satisfy this condition. Since P2 is the side position with much larger space than other positions, we decided to distribute *er* to P2.

The obtained primary vowel allocation is shown in Figure 3. This vowel allocation possesses certain optimal characteristics, inherited from the LPC system. However, one can ask if it is really optimal for the Mandarin Chinese CS system. In the next section, we carry out an experimental study on the lip shape parameter distributions in order to clarify this preliminary allocation.

## Coding of the Mandarin Vowels: Optimization

In this section, we analyze the distributions of the lip parameters for vowels allocated to each hand position based on a new corpus recorded by three native Chinese speakers. In this way, we examine whether the different vowels coded by hand positions are sufficiently separable and distinguishable in our Mandarin Chinese CS system.

**Figure 3.** Preliminary allocation of the 16 Mandarin vowels.

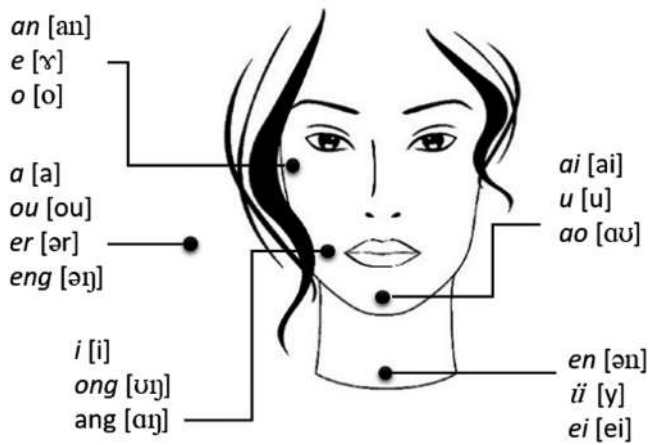

## Database

The database was recorded specifically for this pilot study of the new Mandarin Chinese CS system.[10] It contains three native hearing Chinese speakers (one female and two males). Each of them utters 242 Mandarin words (such as *ting* and *kuang*). Each word is composed of one consonant and one of the 16 Mandarin vowels. These words cover all of the vowel combinations in Mandarin (see Table 3). In Table 4, the vowels' number of occurrences are given. Note that in this corpus, we use words instead of isolated vowels in order to take into account the variability of the syllable context.

The main reason that we use data from hearing speakers is that, according to the original CS system (Cornett, 1967), hand coding is used to distinguish the lip visemes, which are established and analyzed from lip shapes of hearing people.

Temporal segmentation for all vowels is obtained manually using software Praat (Liu, 2018). Lip parameters (i.e., *A* and *B*) are obtained by manually placing several points on the contour of the inner lips. More precisely, the lips width *A* is calculated by the distance of two points $A_1$ and $A_2$, which are the two extremes in the horizontal direction of the inner lip contour (see Figure 4). For calculating the lips height *B*, we define it to be the distance of two points $B_1$ and $B_2$ which are the middle points of the inner lips in the vertical direction (see Figure 4).

## Lip Parameter Distributions for Preliminary Vowel Allocation

After obtaining *A* and *B* parameters of all vowels of the three speakers, we first

**Table 3.** List of 242 Mandarin words (in Pinyin) used for recording the database

| Vowels | Mandarin Words Related to Vowels |
| --- | --- |
| *ei* | ei, gei, shei, wei, lei, fei, mei, tei, bei, nei, hei, dei, pei |
| *ü* | nü, yü, jü, xü, qü, |
| *en* | en, chen, cen, gen, shen, wen, fen, men, ben, nen, sen, ren, hen, ken, zhen, pen |
| *ao* | ao, chao, yao, xiao, cao, gao, shao, lao, mao, tao, bao, nao, sao, rao, zao, kao, zhao, pao |
| *u* | chu, cu, gu, shu, wu, lu, fu, tu, bu, nu, su, ru, hu, zu, ku, zhu, du, pu |
| *ai* | ai, chai, cai, gai, shai, wai, lai, mai, tai, bai, nai, sai, hai, zai, kai, zhai, dai, pai |
| *ang* | chang, yang, cang, xiang, jiang, gang, shang, wang, lang, tang, fang, bang, nang, sang, rang, hang, zang, zhang, dang, pang |
| *ong* | chong, yong, cong, gong, long, tong, nong, song, rong, hong, zong, kong, zhong, dong |
| *i* | chi, yi, xi, ci, ji, mi, ti, bi, ni, si, ri, zi, qi, zhi, di, pi |
| *eng* | cheng, ceng, geng, sheng, weng, leng, feng, meng, teng, beng, neng, seng, reng, heng, keng, zheng, deng, peng |
| *an* | an, chan, yan, xian, can, gan, shan, wan, lan, fan, man, tan, yuan, ban, nan, san, ran, han, zan, kan, zhan, dan, pan |
| *ou* | ou, chou, you, xiou, cou, gou, shou, lou, fou, mou, tou, nou, sou, hou, rou, zou, kou, zhou, dou, pou |
| *a* | a, cha, ya, xia, ca, ga, sha, wa, la, ma, fa, ta, ba, na, sa, ha, za, ka, zha, da, pa |
| *e* | e, che, ye, ce, she, le, me, te, ne, se, re, he, ze, ke, zhe, de |
| *o* | o, wo, fo, bo, po |

*Note.* As for vowel *er*, there is only one.

**Table 4.** Number of occurrences of the 16 Mandarin vowels (i.e., single vowels or compound finals without *i, u, ü* ahead) in the database

| A | O | e | i | u | ü | ai | ei |
| --- | --- | --- | --- | --- | --- | --- | --- |
| 21 | 5 | 16 | 16 | 18 | 5 | 18 | 13 |
| Ang | eng | ong | an | en | er | ou | ao |
| 20 | 18 | 14 | 23 | 16 | 1 | 20 | 18 |

**Figure 4.** Inner lip parameters *A* and *B*.

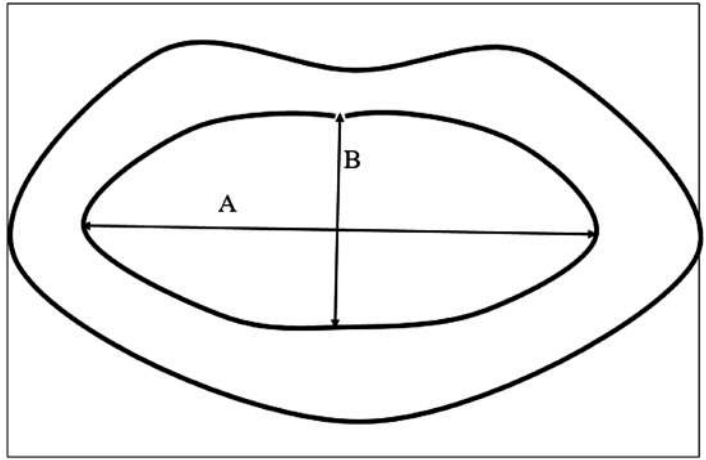

examine the preliminary vowel allocation in *A* and *B* parameter planes. The distributions of these parameters for each speaker are shown in Figures 5, 6, and 7, respectively, with each distribution corresponding to three or four vowels coded by a given hand position.

Now, we analyze the results for the preliminary vowel allocation. Globally, good vowel separation can be observed for all of the speakers (Figures 5–7). Different vowels coded by each hand position are distinguishable in *A* and *B* planes. However, the following specific remarks can be formulated.

1. For Speaker 1 (see Figure 5), some special points are worth mentioning. First, for P1, three vowels *an*, *o,* and *e* are not well-separated. Although, in principle, *e* and *o* are very close but *an* is rather different from them. Indeed, for speakers 2 and 3, *an* is rather separated from *e* and *o*. Secondly, for P3, *ong* is mixed with one part of *i*, caused by its preceding consonants *zh*, *ch*, and *sh*.
2. For Speaker 2 (see Figure 6), *eng* is rather mixed with *a* and *ou* for P2. However, for the other two speakers, this confusion is not marked.
3. For Speaker 3 (see Figure 7), we can see a little confusion between *eng* and *a*, like Speaker 2, but *ou* is totally separated for P2. However, for P5, *en* and *ei* are totally confused. In addition, for P3, *ong* is not mixed with one part of *i* but is very close, like the case for Speaker 1.

In order to objectively measure the separability of the vowels allocated for each hand position, a vowel classification score is calculated. In this experiment, multi-Gaussian models (Bishop, 2006) are used to train and to recognize the vowels in each hand position. More precisely, in each hand position, each vowel is trained by one Gaussian model. Let $x=(A,B)$, which denotes the mean values of *A* and *B* parameters in the target time interval for one vowel. Given any new vowel with lip parameter $x_0=(A_0, B_0)$, we calculate the probability of $x_0$ for each models by Equation 1

$$P(x) = \frac{1}{(2\pi)^{n/2} |\Sigma|^{1/2}} exp\left(-\frac{1}{2}(x-\mu)^T \Sigma^{-1}(x-\mu)\right),$$

where $n = 2$, $\mu$ is the mean value, and $\Sigma$ is the covariance matrix of the Gaussian model. The model with the highest classification probability is the target vowel class.

For example, for position P1, which contains three vowels (*an, o, e*), three Gaussian models are trained for these vowels with *A*, *B* parameters based on our

**Figure 5.** Preliminary lip parameter distributions in *A–B* plane for Speaker 1. Five figures correspond to five hand positions P1 to P5, respectively.

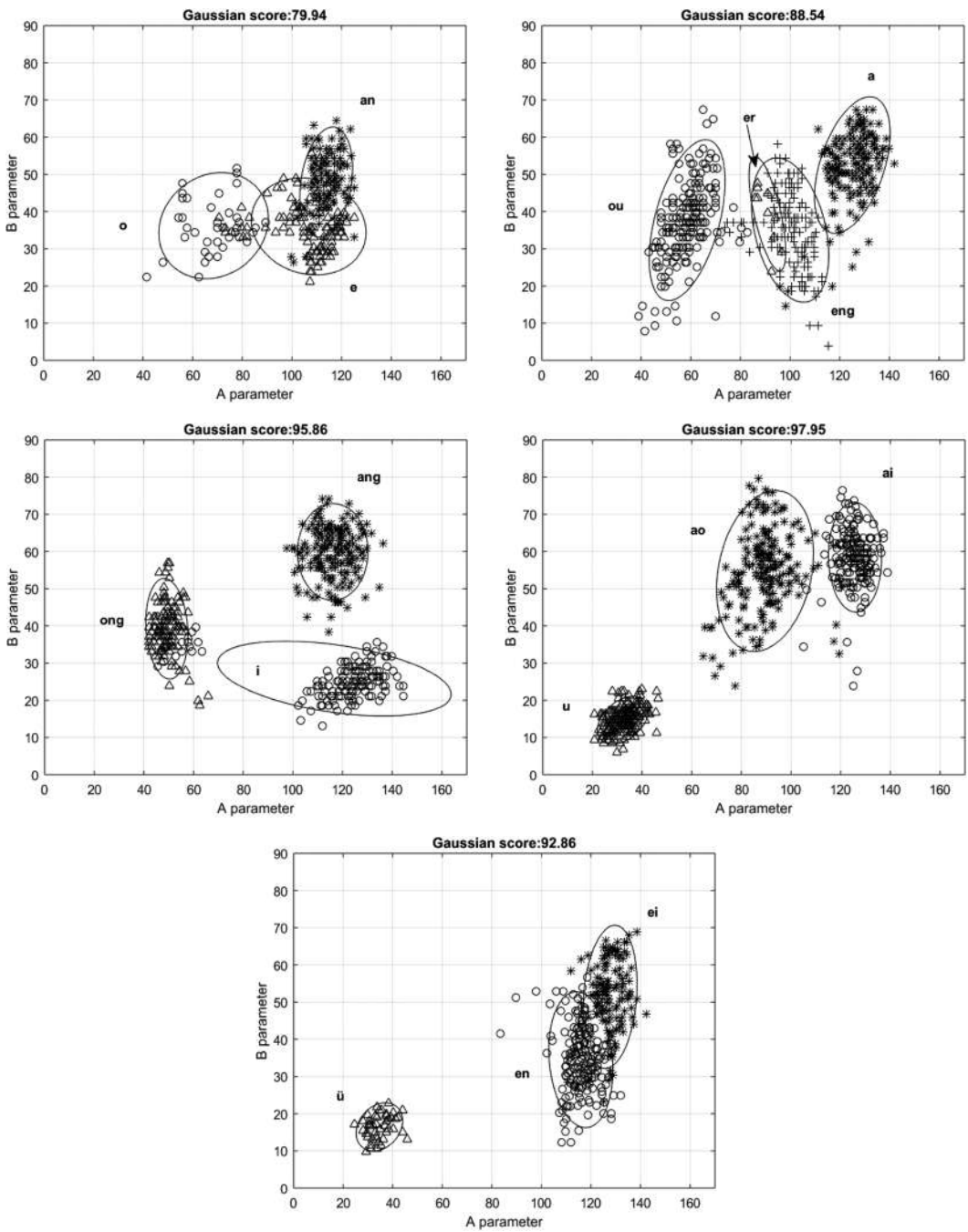

own database for each speaker. Here, 80% of the data is used for training and the remaining 20% is for the test. About 5 images are selected for each vowel. According to Table 2, we can know the number of occurrences for each vowel. For example, concerning vowel *a*, the number of occurrences is 21. Therefore, there are about

**Figure 6.** Preliminary lip parameter distributions in *A*–*B* plane for Speaker 2. Five figures correspond to five hand positions P1 to P5, respectively.

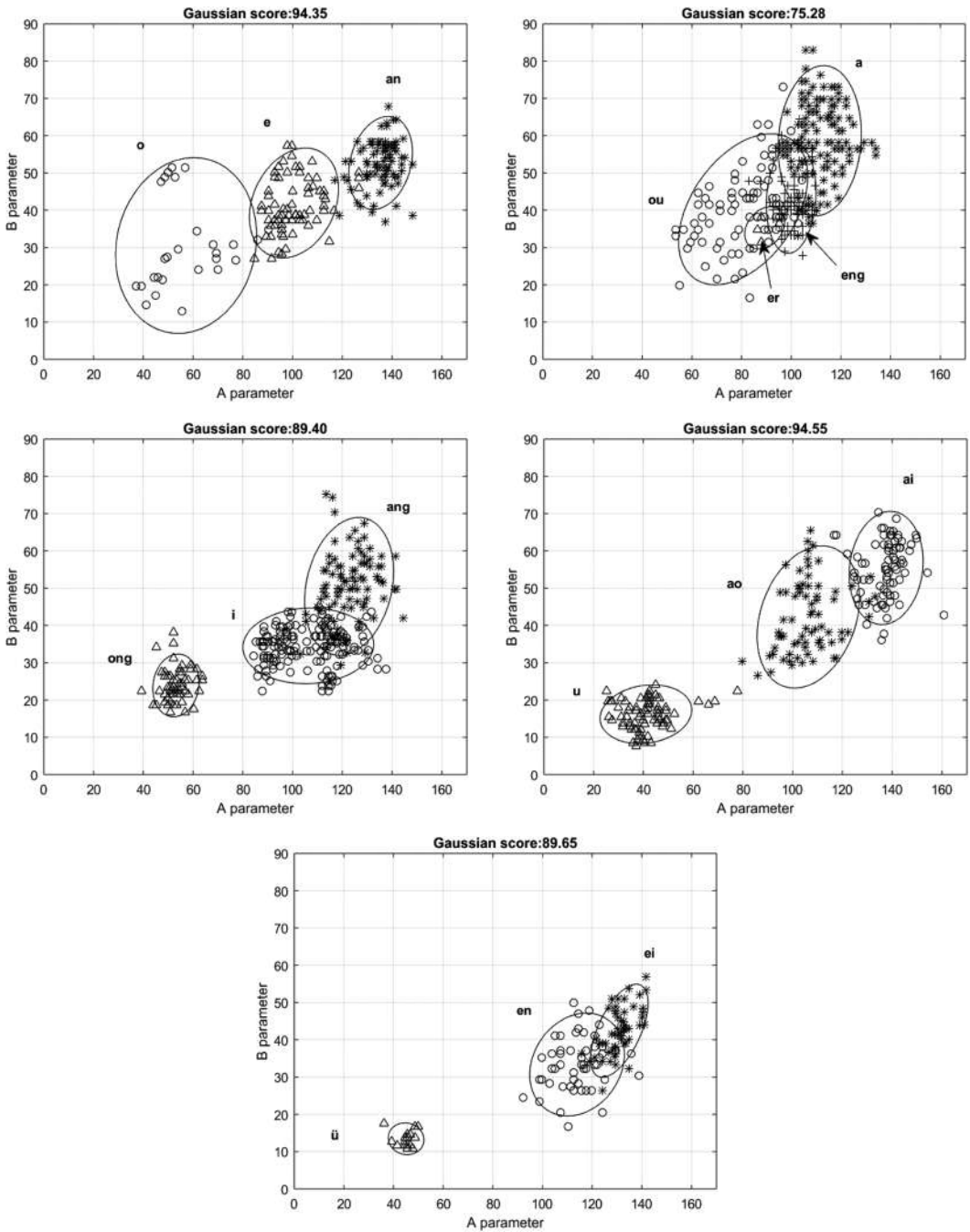

21*5=105 image frames for vowel *a*. In this case, 105*0.8=84 images for vowel *a* are used for training and the rest 21 images are used for the test. The training data of the Gaussian model for vowel *a* is a matrix with size 84*2, where 2 means the *A*, *B* parameters of the inner lip of the image. The result is the average of 100 experiments with the different training and test sets. The standard deviations of all results are less than 1% in Tables 5 and 6.

**Figure 7.** Preliminary lip parameter distributions in *A*–*B* plane for Speaker 3. Five figures correspond to five hand positions P1 to P5, respectively.

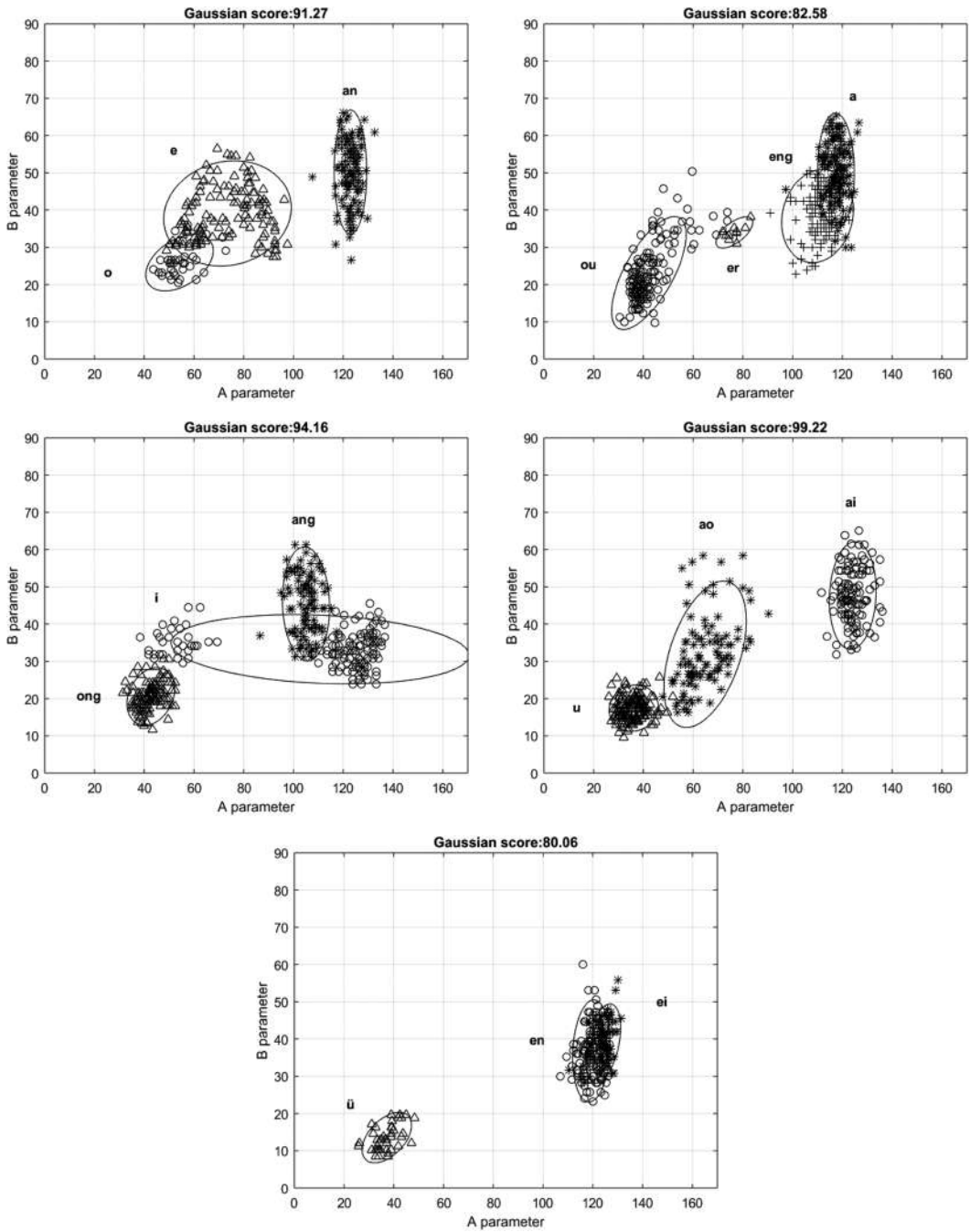

**Table 5.** Gaussian classification scores for different vowels in each hand position (P1–P5), for the preliminary vowel allocation

| Accuracy (%) | P1 | P2 | P3 | P4 | P5 | Average |
|---|---|---|---|---|---|---|
| Speaker 1 | 79.94 | 88.54 | 95.87 | 97.95 | 92.86 | 91.03 |
| Speaker 2 | 94.35 | 75.28 | 89.40 | 94.55 | 89.65 | 86.65 |
| Speaker 3 | 91.27 | 82.58 | 94.16 | 99.22 | 80.06 | 89.46 |

**Table 6.** Gaussian classification scores for different vowels in each hand position (P1–P5), after the optimization

| Accuracy (%) | P1 | P2 | P3 | P4 | P5 | Average |
|---|---|---|---|---|---|---|
| Speaker 1 | 80.01 | 84.65 | 98.63 | 98.12 | 99.01 | 92.08 |
| Speaker 2 | 93.95 | 84.51 | 87.71 | 95.64 | 99.85 | 92.33 |
| Speaker 3 | 92.45 | 81.65 | 95.53 | 98.95 | 95.05 | 92.73 |

Three sets of parameters, in other words, $\Theta_1 = (\mu_1, \Sigma_1)$; $\Theta_2 = (\mu_2, \Sigma_2)$; and $\Theta_3 = (\mu_3, \Sigma_3)$ corresponding to three vowels *an, o,* and *e* are obtained. In the test stage, for a new observation $x_0 = (A_0, B_0)$, using Equation 1, the parameters $\Theta_1$, $\Theta_2$, and $\Theta_3$ corresponding to the maximum probability will be the recognized vowel class.

For each lip parameter distribution, the corresponding classification score obtained by the Gaussian classifier is given (see Table 5). We can see the mean Gaussian classification accuracy for Speaker 1, Speaker 2, and Speaker 3 is 91.03%, 86.65%, and 89.46%, respectively. The accuracy of these classifications objectively indicates the separability of the vowels for each hand position. This means that our primary allocation based on the LPC is rather satisfied.

We then perform an optimization procedure by changing the allocation for several vowels so that the final vowel allocation gives a better separability.

*Optimization*

Based on the previous observations, we make two modifications to reduce the observed confusion in each hand position.

1. Confusion between *ong* and one part of *i* has been observed for Speaker 1, as well as for Speaker 3. In order to avoid it, we propose to exchange the positions of *ong* and *ü* in hand position 5, since *ü* is distributed far from other vowels in the *A* and *B* planes.
2. Vowel *eng* is confused with *a* and *ou* for Speakers 2 and 3, and *en* is also confused with *ei* for Speaker 3. We propose to exchange the positions of *eng* and *en* to resolve this problem.

The results after this optimization are shown in Figures 8, 9, and 10. We can see that the confusion concerning these vowels has been removed.

Furthermore, the Gaussian classification scores in Table 6 confirm the improvement of our optimization. Compared with the result before optimization (Table 5), the average recognition score for each speaker is increased (i.e., Speaker 1: 91.03% vs 92.08%; Speaker 2: 86.65% vs. 92.33%, and Speaker 3: 89.46% vs 92.73% ). These optimized allocations are the ultimate proposed vowel allocations.

**Figure 8.** Lip parameter distributions in *A–B* plane after the optimization, for Speaker 1. Five figures correspond to five hand positions P1 to P5, respectively.

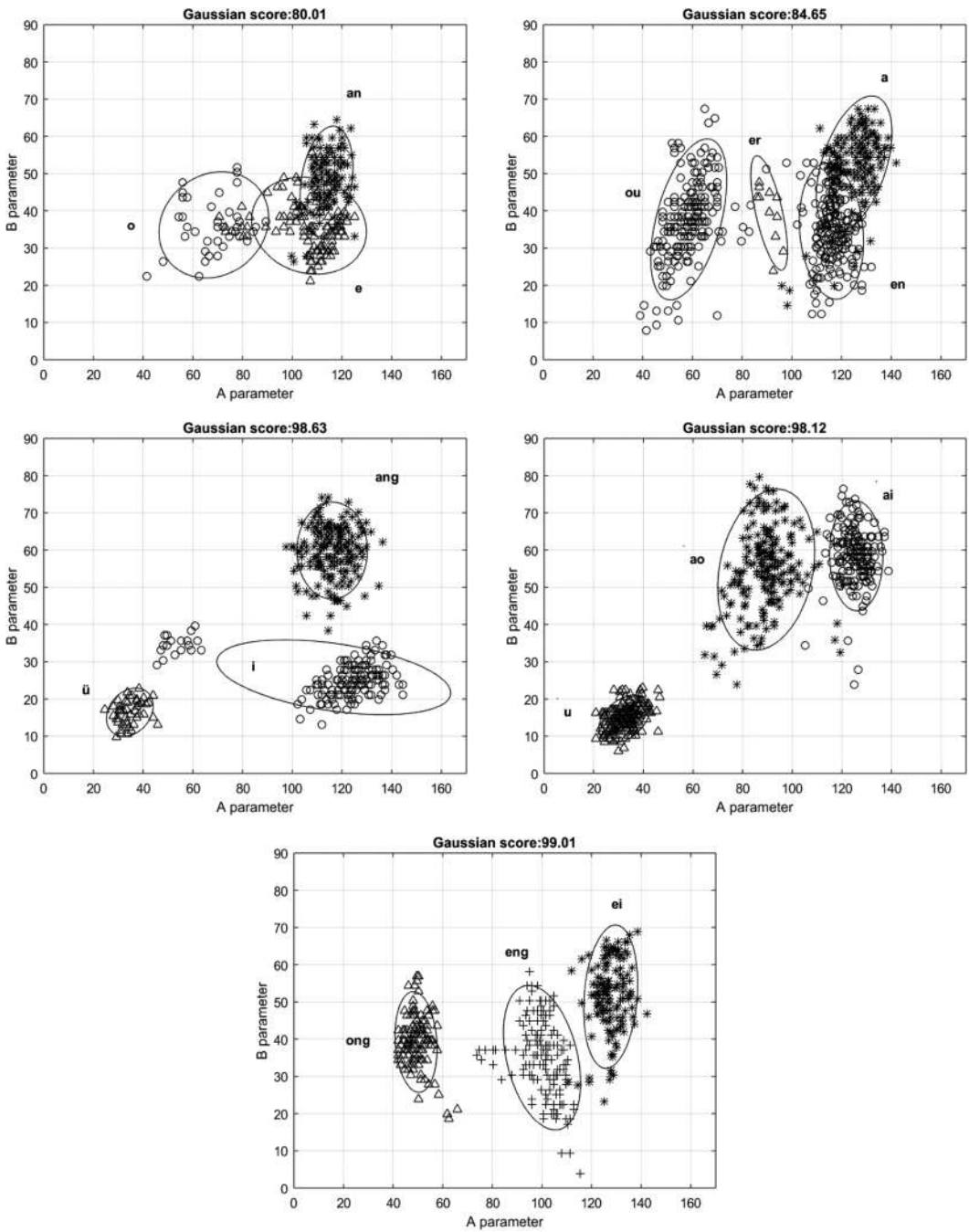

**Figure 9.** Lip parameter distributions in *A–B* plane after the optimization, for Speaker 2. Five figures correspond to five hand positions P1 to P5, respectively.

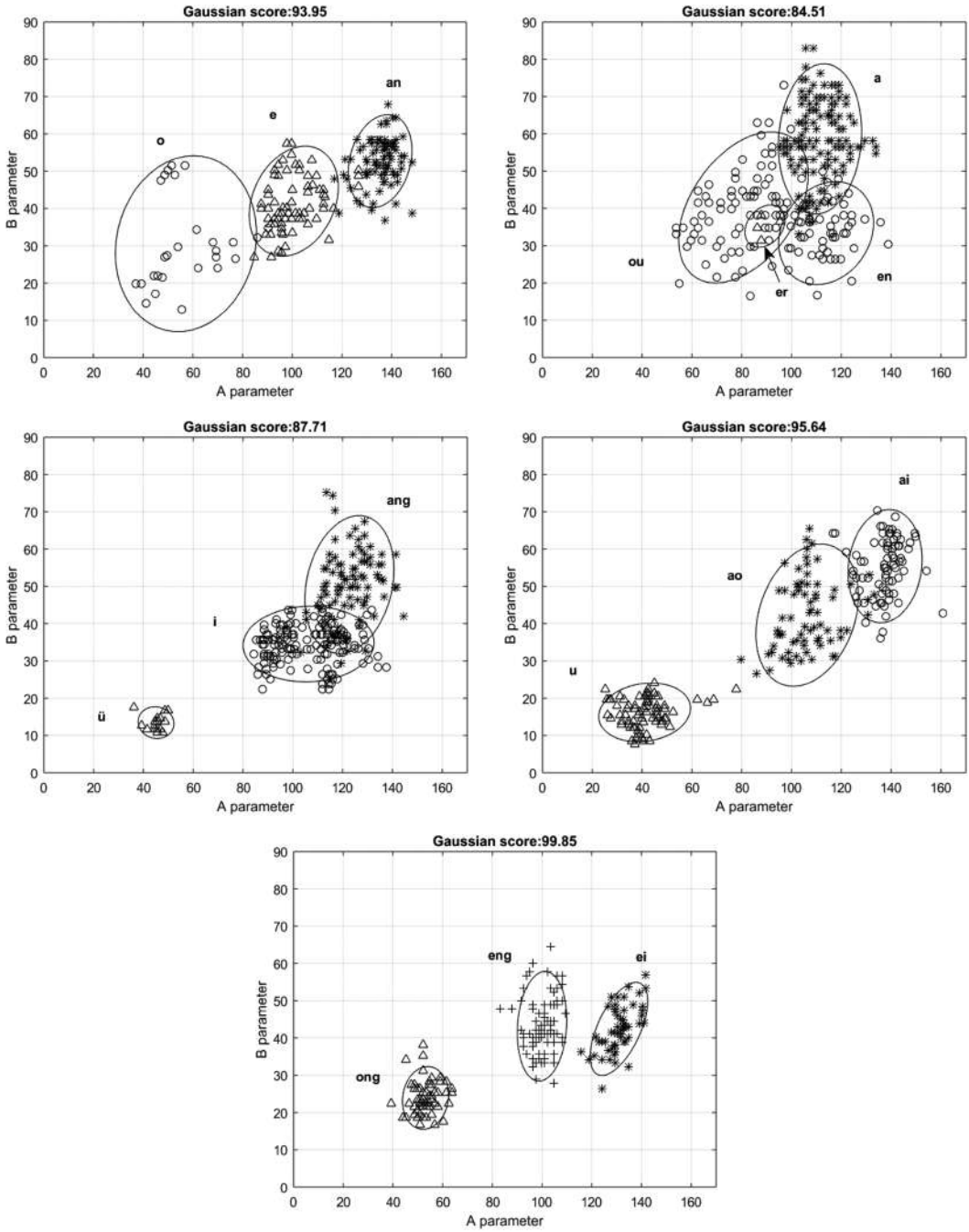

**Figure 10.** Lip parameter distributions in *A–B* plane after the optimization, for Speaker 3. Five figures correspond to five hand positions P1 to P5, respectively.

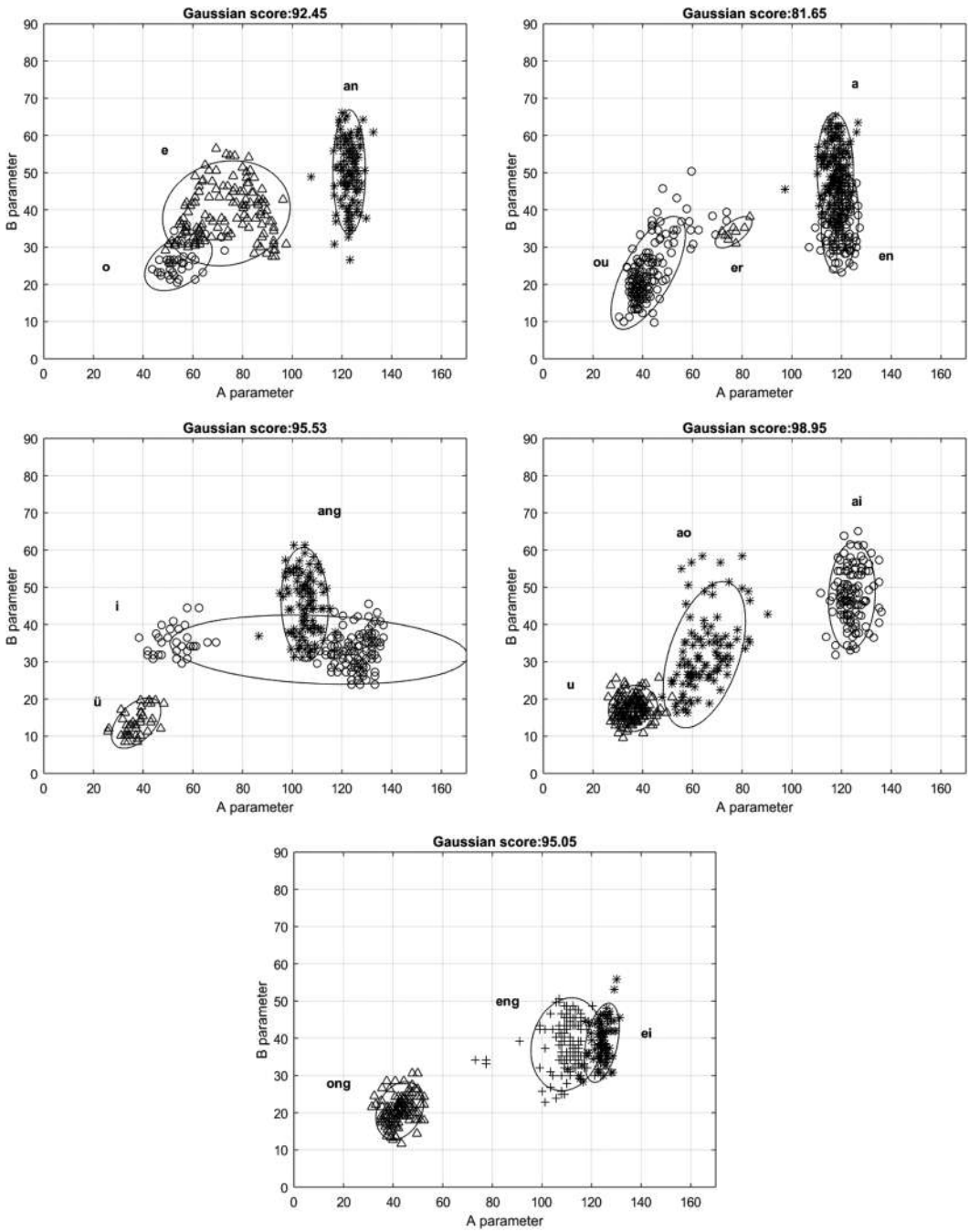

## Coding of Consonants

Considering Mandarin consonants, we adopt the same principle as the coding of Mandarin vowels: The most commonly used consonants are distributed as in the LPC system, which has already been confirmed for optimal performance. Indeed, it has been recommended by Cornett (Cornett, 1994) to maintain the maximum compatibility of the target CS system with other known CS systems. Therefore, in this work, we first try to explore the correspondence between the Mandarin Chinese CS and LPC.

Note that there are 21 Mandarin consonants with 3 semiconsonants (Duanmu, 2007). Therefore, 24 consonants need to be allocated. If one handshape is suitable to indicate three consonants, 8 handshapes are needed to code 24 consonants. Recall that the LPC system is also designed to use 8 handshapes to code 21 consonants.

It should be mentioned that, for consonants, there is a notable difference between Mandarin Pinyin and their corresponding International Phonetic Alphabet (IPA) symbols (Manser et al., 2003). For example, Mandarin consonants *b* and *p* are pronounced as [p, p$^h$] instead of [b, p] (Duanmu, 2007). From the phonetic point of view, this difference is important, especially for the production and perception of these consonants. However, we think this difference is not fundamental when designing a CS system, because the lip shapes of [b, p] and [p, p$^h$] are very similar. In addition, in order to keep the compatibility of Mandarin and *LPC*,[11] it is better to avoid coding Mandarin *b* as the French [p], which brings some confusion for bilingual CS cuers. Consequently, we decided to distribute Mandarin consonants in function of their Pinyin form instead of their phonetic form on the condition that they have similar lip shapes.

**Table 7.** Consonants that can precede a semiconsonant. The correct check symbol means that the combination is possible

| Consonants | Semi-consonants | | |
|---|---|---|---|
| | i [j] | u [w] | ü [ɥ] |
| *b p m f* | ✓ | | |
| *d t n l* | ✓ | ✓ | ✓ |
| *g k h* | | ✓ | |
| *j q x* | ✓ | | ✓ |
| *zh ch sh r z c s* | | ✓ | |

In this way, the following consonants are distributed corresponding to the LPC system: *b*, *p*, *m*, *f*, *d*, *t*, *n*, *l*, *g*, *k*, *sh*, *z*, *s*, and *r*, as well as three semiconsonants [j], [w], and [ɥ]. For the remaining special Mandarin consonants: *h*, *j*, *q*, *x*, *zh*, *ch*, and *c*, their distributions are determined by distinguishing their corresponding lip shapes.

Moreover, in order to mark the handshape change between the consonant and semiconsonants [j], [w], and [ɥ], the consonants that can be combined with [j], [w], [ɥ] should not be placed in the same handshape group. For this purpose, we examine the occurrence of the consonants (Wu & Shih, 2009) that can be combined with the semiconsonants [j], [w], or [ɥ] (see Table 7) to realize a consonant allocation. More precisely, we decided to remove *sh* from the handshape group 6 where there is [w], since there are many combinations between *sh* and [w]. However, it is not possible to totally avoid this situation. In our case, *l* is allocated to the same group 6 as [w] even though combinations between *l* and [w] are possible. This phenomenon can also be found in French CS, and the complementary information from lip patterns can help avoid confusions.

Tone Coding

Mandarin Chinese is a tonal language (Howie, 1976; Gottfried & Suiter, 1997; Wang & Spence, 1999; Xu, 1997; Howie, 1974), which contains four main tones and one neutral tone. The same syllable can be pronounced with different tones, giving different meanings. For example, la (tone 0: the neutral tone); lā (tone 1: high level tone); lá (tone 2: low rising tone); lǎ (tone 3: falling-rising tone); là (tone 4: high falling tone) represent five different words by five different tones.

Wikipedia (https://en.wikipedia.org/wiki/Cued_speech) reports that CS systems of tonal languages, the tone could be indicated by hand inclination and movement. However, we noticed that in the CS coding, cuers' hands will naturally rotate or incline, thus it is not clear if those movements indicate tone or not. Besides, we think that using hand movements for Mandarin tones will be very complicated to achieve because a large number of vowels in Mandarin already require a lot of hand movements.

Therefore, we propose to indicate the Mandarin tones by head movements. In the case of tone 0, the head keeps still, which means no tone indication is needed. In tone 1, the head shifts to the right without any rotation. In tone 2, the head shifts to the up direction. In tone 3, the head moves down and up (a shape V) and in tone 4, it moves down direction (see Fig. 11). Indeed, the tone indications are only needed when there is confusion.[12] For example, the word sōngshù with the 4th tone of shù is totally different with sōngshǔ with the 3rd tone of shǔ.

Based on all these considerations, we propose a complete Mandarin Cued CS system that is shown in Figure 11 (both in Pinyin and IPA). Note that in Mandarin Chinese, the special case of an isolated vowel is coded with handshape P5 (marked by an asterisk [*] in Fig. 11).

Conclusion

In this work, we carried out a pilot study and proposed a novel and efficient Mandarin Chinese CS system that is designed and optimized for the deaf community. In this system, Mandarin compound finals starting with *i*, *u*, and *ü* are coded using the semiconsonants [j], [w], and [ɥ]. By this way, the number of finals that need to be coded by hand positions is reduced from 36 to 16, so that no additional hand slides are used in our system. This is one of the main advantages since reducing hand movements to a minimum level permits CS cuers to expend less energy, increases its efficiency, and allows the CS interlocutor to decode the speech with minimum ambiguity. Besides, we explore similarities between Mandarin and French vowels to build a preliminary Mandarin Chinese CS vowel allocation, which is then optimized by analyzing lip parameter distributions for each hand position based on a new database. This approach permits optimal vowel allocations and guarantees a maximum compatibility with the known LPC system. The consonant coding is also based on the same principle. Finally, we propose to code Mandarin tones by head movements, which reduces the complexity of hand movements.

Limitation of the Current Study

In this work, one of our main objectives is to build a Mandarin Chinese CS system without using any slides for Mandarin compound finals. This makes the proposed Mandarin Chinese CS system efficient since it avoids using lots of hand shifts. However, some cuers who already code with

‎

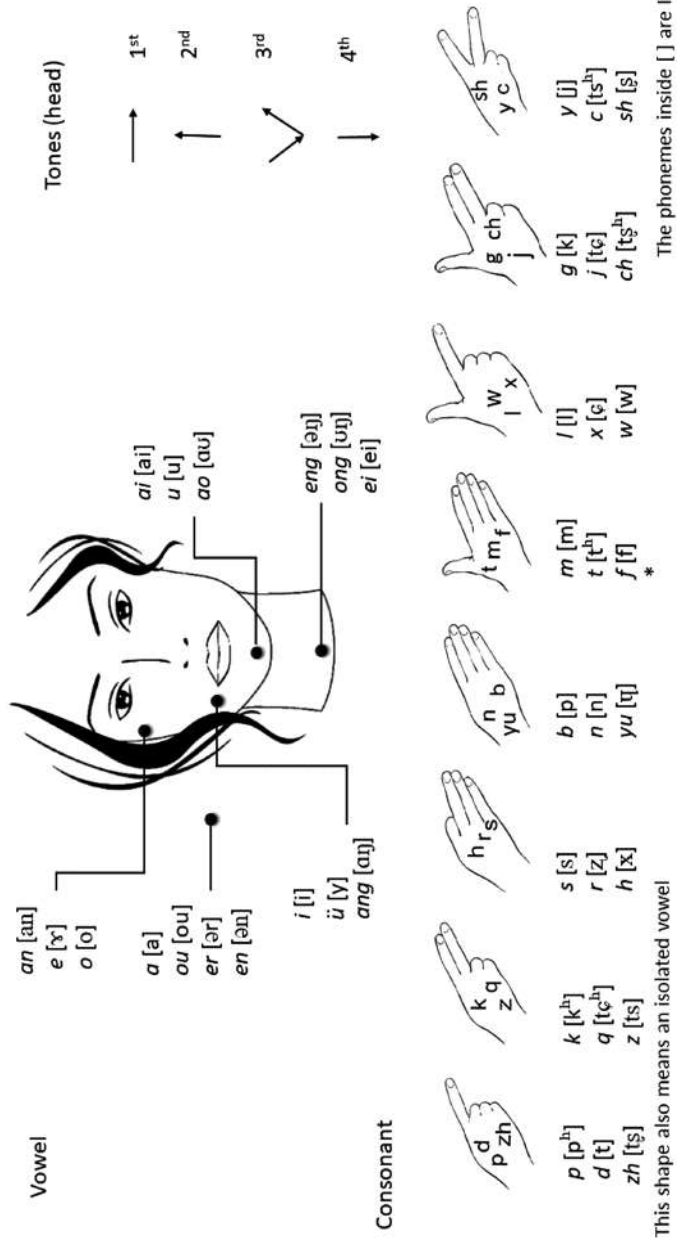

Figure 11. Final chart of the proposed Mandarin Chinese CS system (in Pinyin and IPA).

American/British English CS, might prefer a Mandarin CS system with slides. Therefore, further comparative studies with and without using the slides could be carried out. Cuers' practical feedback would determine if the proposed system is optimal and compatible with other languages of CS.

Future Practice of the Proposed System

In the future, we hope to apply this system in practice to examine its performance, either with hearing families with deaf or hard of hearing children or in deaf schools. For example, a certain number of deaf children can be asked to study the proposed Mandarin Chinese CS system for one semester (i.e., several months). After this semester, we could evaluate their performance by asking them to code 20 Mandarin words or sentences. Moreover, we would ask for their feedback about the proposed system compared with other possible proposals such as using slides for diphthongs.

Acknowledgments

The authors would like to thank the native Chinese speakers for their time spent on the Mandarin Chinese Cued Speech data recording, and professional cuer Hannah Mann for her confirmation and suggestions for the proposed Mandarin Chinese Cued Speech system. Li Liu is a recipient of a scholarship from the Sephora Berrebi Foundation and gratefully acknowledges its support.

Notes

1. French CS is called Langue Francaise Parlée Complétée (LPC)
2. http://www.cuedspeech.org/
3. http://www.cuedspeech.co.uk/
4. http://alpc.asso.fr/
5. http://www.cdpf.org.cn/english/
6. Mandarin Chinese Pinyin is the official Romanization system for standard Chinese in mainland China (Li & Thompson, 1989; Lin, 2001) is in italic type, and the international phonetic symbol form is in [ ].
7. ALPC reports that LPC has been used for about 40 years with very positive feedback from the deaf community (LaSasso et al., 2010).
8. Note that there are only 15 vowels in French with 5 hand positions and about 20 vowels in British/America English with 4 hand positions and 4 hand slides.
9. According to Lin (2007), [ɿ] and [ʅ] are called apical vowels.
10. This dataset will be made publicly available on Zenodo.
11. It is good to keep the consistency and compatibility with the other language CS system, since in this way, CS cuers can switch from one language to another language more easily (Cornett, 1994).
12. In daily communication, the confusion made by tones can be reduced based on the contextual information of the conversation.